\documentclass[prl,showpacs,amsfonts,amsmath,twocolumn,superscriptaddress,a4paper]{revtex4}
\usepackage{array,amsmath,amsfonts,amssymb,dsfont}
\usepackage{natbib}
\usepackage[T1]{fontenc}
\usepackage{ae}
\usepackage{graphicx}

\DeclareMathOperator{\Tr}{Tr}
\begin{document}

\title{Two-Qubit State Tomography using a Joint Dispersive Read-Out}

\affiliation{Department of Physics, ETH Zurich, CH-8093 Zurich, Switzerland}
\author{S.~Filipp}
\email{filipp@phys.ethz.ch}
\affiliation{Department of Physics, ETH Zurich, CH-8093
  Zurich, Switzerland}
\author{P.~Maurer}
\thanks{The first two authors contributed equally to this work.}
\affiliation{Department of Physics, ETH Zurich, CH-8093
  Zurich, Switzerland}
\author{P.~J.~Leek}
\author{M.~Baur}
\author{R.~Bianchetti}
\author{J.~M.~Fink}
\author{M.~G\"oppl}
\author{L.~Steffen}
\affiliation{Department of Physics, ETH Zurich, CH-8093 Zurich, Switzerland}
\author{J.~M.~Gambetta}
\affiliation{Institute for Quantum Computing and Department of Physics
  and Astronomy, University of Waterloo,
  Waterloo, Ontario N2L 3G1, Canada}
\author{A.~Blais}
\affiliation{D\'{e}partement de Physique, Universit\'{e} de Sherbrooke, Sherbrooke, Qu\'{e}bec J1K 2R1, Canada}
\author{A.~Wallraff}
\affiliation{Department of Physics, ETH Zurich, CH-8093 Zurich, Switzerland}


\renewcommand{\i}{{\mathrm i}}
\def\1{\mathchoice{\rm 1\mskip-4.2mu l}{\rm 1\mskip-4.2mu l}{\rm 1\mskip-4.6mu l}{\rm 1\mskip-5.2mu l}}
\newcommand{\ket}[1]{|#1\rangle}
\newcommand{\bra}[1]{\langle #1|}
\newcommand{\braket}[2]{\langle #1|#2\rangle}
\newcommand{\ketbra}[2]{|#1\rangle\langle#2|}
\newcommand{\opelem}[3]{\langle #1|#2|#3\rangle}
\newcommand{\projection}[1]{|#1\rangle\langle#1|}
\newcommand{\scalar}[1]{\langle #1|#1\rangle}
\newcommand{\op}[1]{\hat{#1}}
\newcommand{\vect}[1]{\boldsymbol{#1}}
\newcommand{\id}{\text{id}}

\begin{abstract}
Quantum state tomography is an important tool in quantum information
science for complete characterization of multi-qubit states and
their correlations. Here we report a method to perform a joint
simultaneous read-out of two superconducting qubits dispersively
coupled to the same mode of a microwave transmission line resonator.
The non-linear dependence of the resonator transmission on the qubit
state dependent cavity frequency allows us to extract the full
two-qubit correlations without the need for single shot read-out of
individual qubits. We employ standard tomographic techniques to
reconstruct the density matrix of two-qubit
quantum states.
\end{abstract}

\maketitle

Quantum state tomography allows for the reconstruction of an
a-priori unknown state of a quantum system by measuring a complete
set of observables \cite{Paris2004}. It is an essential tool in the
development of quantum information processing \cite{Nielsen2000} and
has first been used to reconstruct the Wigner-function
\cite{Schleich2001} of a light mode \cite{Smithey1993} by homodyne
measurements, as suggested in a seminal paper by Vogel and Risken
\cite{Vogel1989}. Subsequently, state tomography has been applied to
other systems with a continuous spectrum, for instance, to determine
vibrational states of molecules \cite{Dunn1995}, ions
\cite{Leibfried1996} and atoms \cite{Kurtsiefer1997}. Later,
techniques have been adapted to systems with a discrete spectrum,
for example nuclear spins \cite{Chuang1998}, polarization entangled
photon pairs \cite{White1999}, electronic states of trapped ions
\cite{Roos2004}, states of hybrid atom-photon systems
\cite{Volz2006}, and spin-path entangled single neutrons \cite{Hasegawa2007}.

Recent advances have enabled the coherent
control of individual quantum two-level systems embedded in a solid-state environment. Numerous experiments have been performed 
with superconducting quantum devices \cite{Clarke2008}, manifesting
the rapid progress and the promising future of this approach to
quantum information processing. In particular, the strong coupling
of superconducting qubits to a coplanar waveguide resonator can be
exploited to perform cavity quantum electrodynamics (QED)
experiments on a chip \cite{Wallraff2004b,Blais2004,Schoelkopf2008} in an
architecture known as circuit QED. The high level of control over the
dynamics of this coupled quantum system has been demonstrated,
e.~g., in \cite{Leek2007,Hofheinz2008}. State
tomographic methods have already been used in superconducting
circuits to verify the entanglement between two phase qubits
\cite{Steffen2006a}. There, the state is determined for each
individual qubit with single-shot read-out such that two-qubit
correlations can be evaluated by correlating the single measurement
outcomes. In contrast, in this letter we extract two-qubit
correlations from the simultaneous averaged measurement of two
qubits dispersively coupled to a common resonator. This possibility
has also been pointed out in Ref.~\onlinecite{Majer2007}.

\begin{figure}[!b]
  \centering
    \includegraphics[width=86mm]{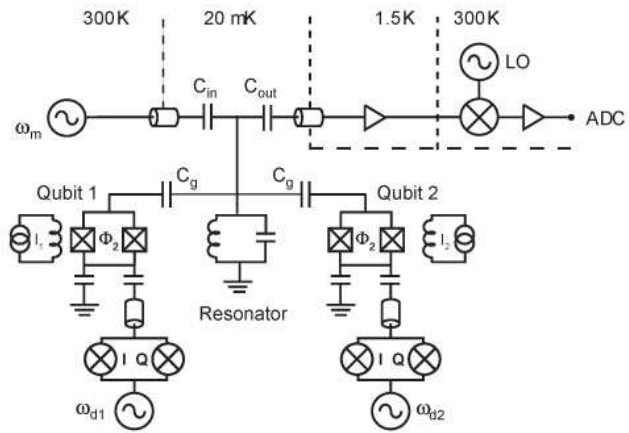}
\caption{Schematic of the experimental setup with two qubits coupled
via the capacitances $\rm{C}_{\rm{g}}$ to a microwave resonator
operated at a temperature of about $20~\rm{mK}$. The transition
frequencies of the qubits are adjusted by external fluxes $\Phi_1$
and $\Phi_2$. The resonator-qubit system is probed through the input
and output capacitances $\rm{C}_{\rm{in}}$ and $\rm{C}_{\rm{out}}$
by a microwave signal at frequency $\omega_{m}$. Additionally, local
control of the qubits is implemented by capacitively coupled signals
$\omega_{d1}$ and $\omega_{d2}$, which are phase and amplitude modulated using
IQ mixers. The output signal is detected in a homodyne measurement
at room temperature.} \label{fig:setup}
\end{figure}

In the setup shown in Fig.~\ref{fig:setup}, two superconducting
qubits are coupled to a transmission line resonator operating in the
microwave regime \cite{Majer2007}.
Due to the large dipole moment of the qubits and the large vacuum
field of the resonator the strong coupling regime with $g_{1,2}\gg
\kappa,\gamma_1$ is reached. $g_1/2\pi \approx g_2/2\pi = 133~\rm{MHz}$
denotes the similar coupling strenghts of both qubits and
$\kappa/2\pi\approx 1.65~\rm{MHz}$,
$\gamma_1/2\pi \approx 0.25~\rm{MHz}$ the photon and the qubit decay
rates, respectively. The qubits are realized as transmons
\cite{Koch2007}, a variant of a split Cooper pair box
\cite{Bouchiat1998} with exponentially suppressed sensitivy to 1/f
charge noise \cite{Schreier2008}. The transition frequencies
$\omega_{aj}$ ($j=1,2$) of the qubits are tuned separately by
external flux bias coils. Both qubits can be addressed individually
through local gate-lines using amplitude and phase modulated
microwaves at frequencies $\omega_{d1}$ and $\omega_{d2}$. Read-out is accomplished by measuring the transmission of microwaves
applied to the resonator input at frequency $\omega_m$
close to the fundamental resonator mode $\omega_r$.  
At large detunings $\Delta_{j} \equiv \omega_{aj}-\omega_r$ of both
qubits from the resonator, the dispersive qubit-resonator interaction
gives rise to a qubit state dependent shift of the resonator
frequency. In this dispersive limit and in a frame rotating at
$\omega_m$ the relevant Hamiltonian reads \cite{Blais2007}
\begin{eqnarray}
  \label{eq:hdisp}
  H &=& \hbar\big(\Delta_{rm} + \chi_1\op{\sigma}_{z1}+\chi_2
        \op{\sigma}_{z2}\big)\op{a}^\dagger \op{a}\\\nonumber
        &&+\frac{\hbar}{2}\sum_{j=1,2}\left(\omega_{aj}+\chi_j\right)\op{\sigma}_{zj} 
        + \hbar\epsilon(t) (\op{a}^\dagger + \op{a}),
\end{eqnarray}
where $\Delta_{rm}\equiv \omega_r - \omega_m$ is the detuning of the
measurement drive from the resonator frequency. The coefficients
$\chi_{1,2}$ are determined by the detuning $\Delta_{1,2}$, the
coupling strength $g_{1,2}$ and the design parameters of the qubit
\cite{Koch2007}. The last term in Eq.~(\ref{eq:hdisp}) models the
measurement drive with amplitude $\epsilon(t)$.

The operator $\op{\chi}\equiv \chi_1\op{\sigma}_{z1}+\chi_2
        \op{\sigma}_{z2} $, which describes the dispersive shift of
the resonator frequency, is linear in both qubit states. It does not
contain two-qubit terms like $\op{\sigma}_{z1}\op{\sigma}_{z2}$
from which information about the qubit-qubit correlations could be
obtained. However, in circuit QED instead of measuring frequency
shifts directly, we record quadrature amplitudes of microwave
transmission through the resonator which depend nonlinearly on these
shifts. The average values of the field quadratures $\langle
\op{I}(t)\rangle = \Tr[\op{\rho}(t)(\op{a}^\dagger +
\op{a})]$ and $\langle \op{Q} \rangle =
\i\Tr[\op{\rho}(t)(\op{a}^\dagger - \op{a})]$ are
determined from the amplified voltage signal at the resonator output
in a homodyne measurement. Here,  $\op{\rho}(t)  = U_m(t)
\op{\rho}(0)U_m(t)^\dagger$  denotes the
time evolved state of both resonator and qubit under measurement. In
the dispersive approximation we can safely
assume this state to be separable before the measurement, which is
taken to start at time
$t_m$,  $\op{\rho}(t_m) = \ketbra{0}{0}\otimes
\op{\rho}_q(t_m)$. 
Using these expressions, we find $\langle \op{Q}(t) \rangle = \i
\Tr_q[\op{\rho}_q(t_m)\bra{0}\op{U}^\dagger_m(t)
(\op{a}^\dagger - \op{a}) \op{U}_m(t)\ket{0}]$ (and similarly
for $\langle \op{I}(t) \rangle$), where $\Tr_q$
denotes the partial trace over the qubit. This expression is
evaluated using the input-ouput formalism \cite{Gardiner1992} including
cavity decay $\kappa$. In the steady-state, this yields $\langle
\op{I} \rangle_s, \langle Q \rangle_s = -\epsilon \Tr_q
[\rho_q(t_m)\op{M}_{I,Q}]$ with
  \begin{eqnarray}
  \label{eq:it}
    \op{M}_I &=&  \frac{2(\Delta_{rm} + \op{\chi})}{(\Delta_{rm} +\op{\chi})^2 +
      (\kappa/2)^2},\\%\text{ and}
    \op{M}_Q &=&\frac{\kappa}{(\Delta_{rm} + \op{\chi})^2 +
      (\kappa/2)^2}.
  \end{eqnarray}
We note that the measurement operators are nonlinear functions
of $\op{\chi}$. Thus, $\op{M}_{I,Q}$ comprises in general also
two-qubit correlation terms
proportional to $\sigma_{z1} \sigma_{z2}$, which allow to
reconstruct the full two-qubit state.

\begin{figure}[!t]
  \centering
  \includegraphics[width=86mm]{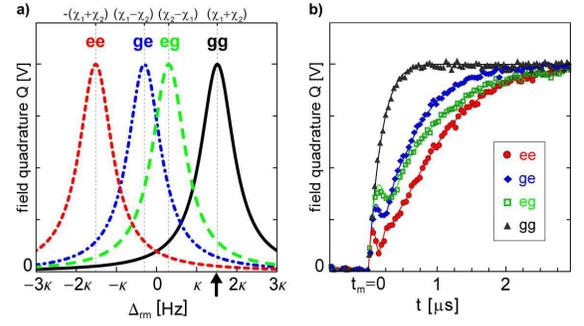}
%statetomography/IQMeasurement.nb
\caption{(a) $Q$-quadrature of the resonator field for the qubits in
states $gg$, $eg$, $ge$ and $ee$ as a function of the detuning
$\Delta_{rm}$. Tomography measurements have been performed at
$\Delta_{rm}=(\chi_1+\chi_2)$ indicated by an arrow. (b) Measured
(data points) time evolution of the $Q$-quadrature for the indicated
initial states compared to numerically calculated responses (solid
lines). All parameters have been determined in independent
measurements.}
  \label{fig:phaseshift}
\end{figure}

In our experiments the phase of the measurement microwave at
frequency $\Delta_{rm} = (\chi_1+\chi_2)$ is adjusted such that the
$Q$-quadrature of the transmitted signal carries most of the signal
when both qubits are in the ground state. The corresponding
measurement operator can be expressed as
\begin{equation}
\label{eq:M}
\op{M} =\frac{1}{4}\left( \beta_{00} \op{\id} +  \beta_{10} \op{\sigma}_{z1} + \beta_{01} \op{\sigma}_{z2} + \beta_{11} \op{\sigma}_{z1} \op{\sigma}_{z2}\right),
\end{equation}
where $\beta_{ij} = \alpha_{\text{\scriptsize -\,-}} + (-1)^{j}
\alpha_{\text{\scriptsize -+}} + (-1)^i \alpha_{\text{\scriptsize
+\,-}} + (-1)^{i+j}\alpha_{\text{\scriptsize ++}}$, with the
coefficients
\begin{equation}
\label{eq:alpha}
\alpha_{\text{\scriptsize $\pm\pm$}} = -\epsilon\kappa \{(\kappa/2)^2 + (\Delta_{rm} \pm \chi_1 \pm
  \chi_2)^2\}^{-1/2}
\end{equation}
representing the qubit state dependent $Q$-quadrature amplitudes of
the resonator field in the steady-state limit and for infinite
qubit-lifetime (Fig.~\ref{fig:phaseshift}(a)).

Since we operate in a regime, where the qubit relaxation cannot be
neglected, the steady-state expression is of limited practical use. The
decay of a qubit to its ground state changes the resonance frequency
of the resonator and consequently limits the read-out time to
$\sim 1/\gamma_1$. A typical averaged time-trace of the resonator
response for pulsed measurements is shown in
Fig.~\ref{fig:phaseshift}(b), similar to the data presented in
\cite{Majer2007}. The qubits are prepared initially
in the states $\ket{ee}$, $\ket{eg}$, $\ket{ge}$ and $\ket{gg}$,
respectively, using the local gate lines. The time dependence of the
measurement signal is determined by the rise time of the resonator
and the decay time of the qubits. It is in excellent agreement with
calculations  (solid lines in Fig.~\ref{fig:phaseshift}(b)) of the
dynamics of the dispersive Jaynes-Cummings Hamiltonian \cite{Blais2007,Bianchetti2008}.

Due to the quantum non-demolition nature of the measurement
\cite{Blais2004}, $\op{M}$
remains diagonal in the instantaneous qubit eigenbasis during the measurement
process. Therefore, a suitable realistic measurement operator $\op{M}'$ can be
defined by replacing the $\alpha_{\pm\pm}$ in Eq.~(\ref{eq:alpha})
with the integrated signal from $t_m$ to the final time
$T$, $\alpha_{\pm\pm}' = 1/N \int_{t_m}^T
(\langle \op{M}(t) \rangle_{\pm\pm} - \langle\op{M}(t)
\rangle_{\text{\scriptsize -\,-}}) dt$ with the ground state
response $\langle\op{M}(t)\rangle_{\text{\scriptsize -\,-}}$
subtracted. The normalization
constant $N$ is chosen such that $\alpha_{\text{\scriptsize
    +\,-}}'=1$.

To reconstruct the combined state $\op{\rho}_q$ of both qubits, a suitable set
of measurements has to be found to determine unambiguously the 16
coefficients $r_{ij}$ of the density matrix $\op{\rho}_q =
\sum_{i,j=0}^3 r_{ij}\, \op{\sigma}_{i} \otimes \op{\sigma}_{j}$
with the identity $\op{\sigma}_0=\op{\id}$  and
$\{\op{\sigma}_{1},\op{\sigma}_{2},\op{\sigma}_{3}\} =
\{\op{\sigma}_{x},\op{\sigma}_{y},\op{\sigma}_{z}\}$.
Such a complete set of measurements is constructed by applying
appropriate single qubit rotations $\op{U}_k \in SU(2)\otimes SU(2)$ before
the measurement in order to measure the expectation values $\langle \op{M}_k\rangle = \Tr[\op{M} \op{U}_k\op{\rho}_q
  \op{U}_k^\dagger] = \Tr[\op{U}_k^\dagger \op{M}
  \op{U}_k\op{\rho}_q]$. The latter equality defines the set of
measurement operators $\op{M}_k\equiv
\op{U}_k^\dagger \op{M} U_k$. This illustrates again that a
measurement operator $\op{M}$ involving non-trivial two-qubit terms
$\sigma_{i1} \sigma_{j2}$ is necessary for state
tomography. In fact, single-qubit operations $U_k = U_{k1} \otimes U_{k2}$ alone cannot be used to generate correlation terms
  since $U_k^\dagger (\op{\id} \otimes \sigma_{z}) U_k = \op{\id} \otimes
  (U_{k2}^\dagger\sigma_{z} U_{k2})$, for instance.
As
  $\Tr[(\sigma_{k}\otimes\sigma_{l})(\sigma_{m}\otimes\sigma_{n})]=\delta_{km}\delta_{ln}$, some coefficients $r_{ij}$ of the density matrix $\op{\rho}_q$ would not be determined in an averaged measurement.

To identify the  coefficients $r_{ij}$  we
perform 16 linearly-independent measurements. The condition for the
completeness of the set of tomographic measurements is the
non-singularity of the matrix $A$ defined by the relation $\langle \op{M}_k\rangle =
\sum_{l=0}^{15} A_{kl} r_l$ between the the expectation values
$\langle \op{M}_k\rangle$ and the coefficients of the density matrix $r_l$ with $l\equiv
i+4j$. This condition is only violated if one of the coefficients
$\beta_{ij}$ of $\op{M}$ in Eq.~(\ref{eq:M}) vanishes. For instance,
$\beta_{01}=\beta_{10}=0$ for $\Delta_{rm}=0$, which reflects the
fact that we cannot distinguish two identical qubits due to symmetry
reasons as apparent from Fig.~\ref{fig:phaseshift}(a).

\begin{figure}[!t]
  \centering
  \includegraphics[width=86mm]{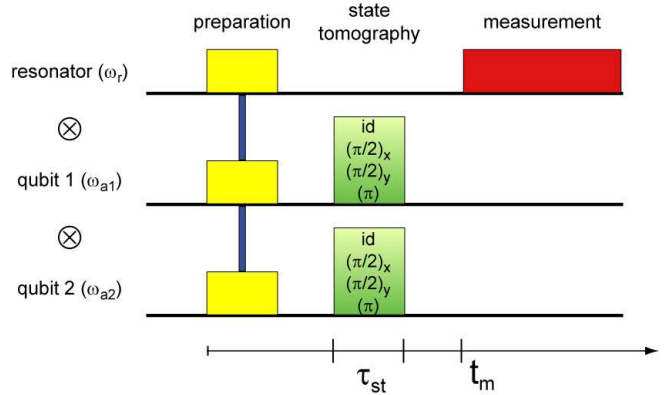}
  \caption{Pulse scheme for state tomography, see text.}
  \label{fig:pulsescheme}
\end{figure}

\begin{figure*}
  \centering
%Peter_32.nb
%PhiMinusBellState-MaxLike.nb
\includegraphics[width=172mm]{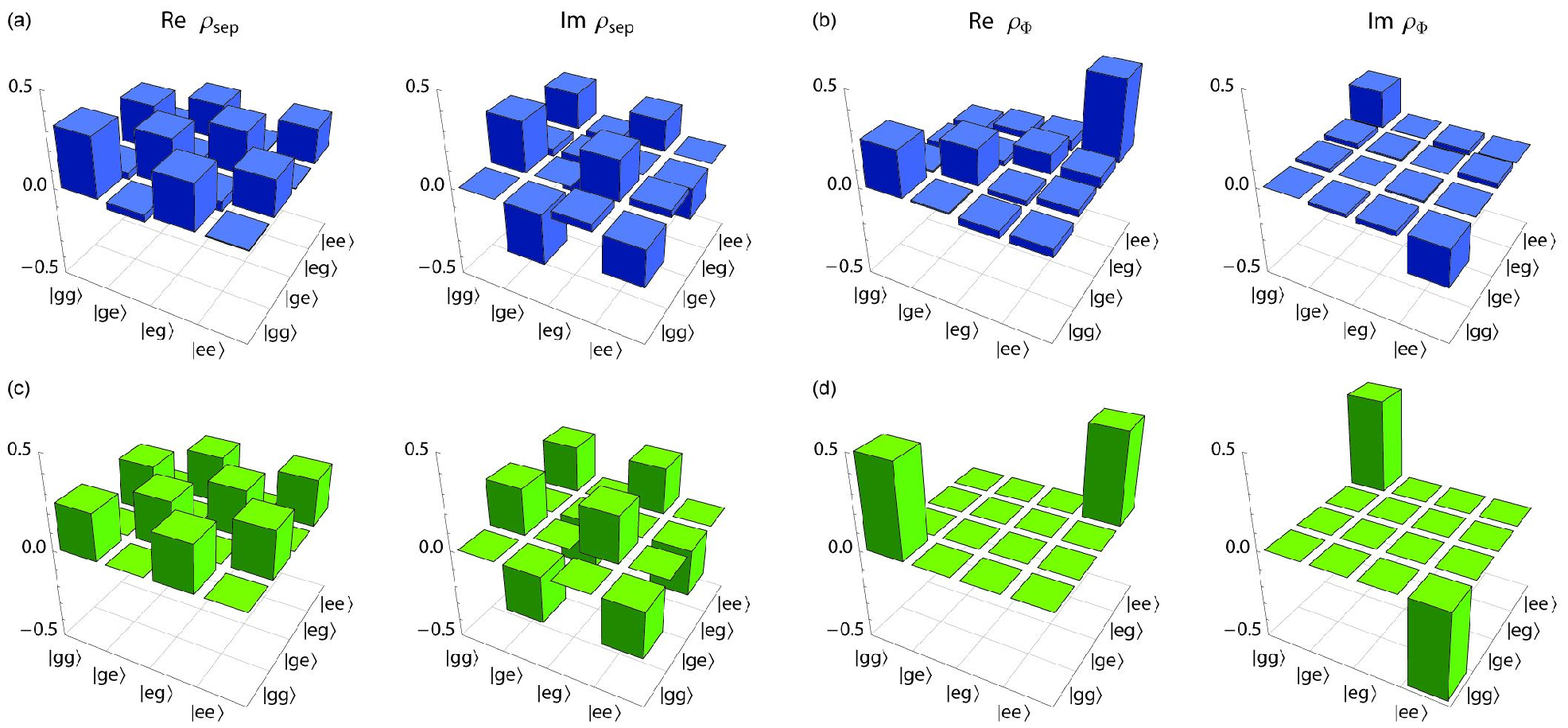}
\caption{Real and imaginary part of reconstruced density matrices of
(a) the product state
$\ket{\Psi_{\rm{sep}}}=1/\sqrt{2}\left(\ket{g}+\ket{e}\right)\otimes
1/\sqrt{2}\left(\ket{g}+\i\ket{e}\right)$  and (b) the Bell state
$\ket{\Phi}=1/\sqrt{2}\left(\ket{g}\otimes\ket{g}-\i\ket{e}\otimes\ket{e}\right)$. Ideal tomograms are shown in (c,d).}
  \label{fig:statetomprod}
\end{figure*}

Our pulse scheme for the state tomography is shown in
Fig.~\ref{fig:pulsescheme}. The transition frequencies of the qubits
are adjusted to $\omega_{a1}/2\pi = 4.5~\rm{GHz}$ and
$\omega_{a2}/2\pi = 4.85~\rm{GHz}$. At this detuning from the
resonator frequency $\omega_r/2\pi = 6.442~\rm{GHz}$ the cavity
pulls are $\chi_1 = - 1~\rm{MHz}$ and $\chi_2=-1.5~\rm{MHz}$
\cite{Koch2007}. First, a given two-qubit state is prepared. Then a
complete set of tomography measurements is formed by applying the
combination of $\{(\pi/2)_x,\,(\pi/2)_y,(\pi),\id\}$ pulses to both
qubits over their individual gate lines using amplitude and phase
controlled microwave signals. The wanted rotation angles are realized
with an accuracy better than $4^\circ$. Finally, the measurement drive
is applied at $\omega_m = 6.445~\rm{GHz}$ corresponding to the
maximum transmission frequency of the resonator with both qubits in
the ground state.
%AnalysisOfTheCalibration.nb
%
To determine the measurement operator $\op{M}'$, $\pi$-pulses are
alternately applied to both qubits to yield signals as shown in
Fig.~\ref{fig:phaseshift}(b). From this data the coefficients
$(\beta_{00}',\beta_{01}',\beta_{10}',\beta_{11}')=(0.8,-0.3,-0.4,-0.1)$
of $\op{M}'$ are deduced, where the non-vanishing $\beta_{11}'$ allows for
a measurement of arbitrary quantum states.
As an example of this state reconstruction,
in Fig.~\ref{fig:statetomprod}(a) the extracted density matrix
$\op{\rho}_{q}$ of the product state
$\ket{\Psi_{\rm{sep}}}=1/\sqrt{2}\left(\ket{g}+\ket{e}\right)\otimes
1/\sqrt{2}\left(\ket{g}+\i\ket{e}\right)$ is shown.
 In
Fig.~\ref{fig:statetomprod}(b) the Bell state $\ket{\Phi} =
1/\sqrt{2}\left(\ket{g}\otimes\ket{g}-\i\ket{e}\otimes\ket{e}\right)$
prepared by a sequence of sideband pulses \cite{Wallraff2007,Blais2007,Leek2008} is
reconstructed. $6.6\times 10^4$ and $6.6\times 10^5$ records have been
averaged, respectively, for each of the 16
tomographic measurement pulses to determine the expectation values
$\langle \op{M}_k'\rangle$ for the two states.
The corresponding ideal state tomograms
 are depicted in Fig.~\ref{fig:statetomprod}(c) and
(d). To avoid unphysical, non positive-semidefinite, density matrices
originating from statistical uncertainties, all tomography data has
been processed by a maximum likelihood method
\cite{Hradil1997,James2001}. The corresponding fidelities
$\mathcal{F}_{\psi} \equiv (\bra{\psi}\op{\rho}_{q}\ket{\psi})^{1/2}$
are $\mathcal{F}_{\rm{sep}} = 95\%$ and $\mathcal{F}_{\Phi^-}=74\%$. These
results are in
close agreement with theoretically expected fidelities when taking
finite photon and qubit-lifetimes into account. As a result, the loss in fidelity and concurrence of the Bell state are not due to measurement errors but to the long preparation sequence \cite{Leek2008}.

In conclusion, we have presented a method to jointly and
simultaneously read-out the full quantum state of two qubits
dispersively coupled to a microwave resonator. In a measurement of
the field quadrature amplitudes of microwaves transmitted through
the resonator each photon carries information about the state of
both qubits. In this way the two-qubit correlations can be extracted
from an averaged measurement of the transmission amplitude without
the need for single shot or single qubit read-out. This method can
also be extended to multi-qubit systems coupled to the same
resonator mode.

This work was supported by Swiss National Science Foundation (SNF)
and ETH Zurich. P.~J.~L. was supported by the EC with a MC-EIF,  J.~M.~G. by CIFAR, MITACS and
ORDCF and A.~B. by NSERC and CIFAR.

\end{document}